\documentclass[a4paper, 10pt, conference]{ieeeconf}      % Use this line for a4 paper

\IEEEoverridecommandlockouts                              % This command is only needed if 
\usepackage{graphicx} % for pdf, bitmapped graphics files
\usepackage{amsmath} % assumes amsmath package installed
\usepackage{amssymb}  % assumes amsmath package installed
\usepackage[style=numeric,sorting=none,doi=false,url=false]{biblatex}
\usepackage{xcolor}
\usepackage{hyperref}
\usepackage{siunitx}

\newcommand{\mbf}[1]{\mathbf{#1}}
\newcommand{\argmin}{\mathop{\arg\min}}
\newcommand{\argmax}{\mathop{\arg\max}}

\title{\LARGE \bf
Tram Positioning with Map-Enabled GNSS Data Reconciliation*
(preprint)}

\author{Jakub Kašpar$^{1}$, Vít Fanta${}^{1}$ and Vladimír Havlena$^{1}$  % <-this % stops a space
\thanks{*This research was supported by the Technological Agency of the Czech Republic program National Competence Centres II, project \# TN02000054 {\em Božek Vehicle Engineering National Center of Competence} (BOVENAC) and by the Grant Agency of the Czech Technical University in Prague, grant No.
SGS22\slash166\slash OHK3\slash 3T/13.}% <-this % stops a space
\thanks{$^{1}$Authors are with the Faculty of Electrical Engineering, Department of Control Engineering, Czech Technical University in Prague {\tt\footnotesize \{kaspaj30, fantavit, havlena\}@fel.cvut.cz}}%
}

\begin{document}

\maketitle
\thispagestyle{empty}
\pagestyle{empty}

%%%%%%%%%%%%%%%%%%%%%%%%%%%%%%%%%%%%%%%%%%%%%%%%%%%%%%%%%%%%%%%%%%%%%%%%%%%%%%%%
\begin{abstract}

This paper presents an approach to tackle
the problem of tram localization through utilizing a custom
processing of Global Navigation Satellite System (GNSS) observables and the track map.
The method is motivated by suboptimal performance
in dense urban environments where the direct line of sight to GNSS satellites
is often obscured which leads to multipath propagation of GNSS signals. The presented concept is based upon the iterated 
extended Kalman filter (IEKF) and has linear complexity (with respect to the number
of GNSS measurements) as opposed to some other techniques mitigating the multipath
signal propagation. The technique is demonstrated both on a simulated example
and real data. The root-mean-squared errors from the simulated ground truth positions
show that the presented solution is able to improve performance compared to a
baseline localization approach. Similar result is  achieved for the 
experiment with real data, while treating orthogonal projections onto the 
tram track as the true position, which is unavailable in the realistic scenario.
This proof-of-concept shows results which may be further improved with
implementation of a bank-of-models method or $\chi^2$-based rejection of 
outlying GNSS pseudorange measurements.

\end{abstract}

%%%%%%%%%%%%%%%%%%%%%%%%%%%%%%%%%%%%%%%%%%%%%%%%%%%%%%%%%%%%%%%%%%%%%%%%%%%%%%%%
\section{INTRODUCTION}
Reliable position estimation is a prerequisite for many advanced driver assistance and anti collision systems in trams. GNSS reception in urban canyon environment is frequently disturbed by {\em multi-path} or {\em non-line-of-sight} (NLOS) propagation of satellite signals resulting in significant position bias. Typically, several satellites from an NLOS segment of the sky are affected.

Classical {\em receiver autonomous integrity monitoring} (RAIM) algorithms developed for aerospace applications rely on high data redundancy and assume only a single fault detection \cite{parkinsonGlobalPositioningSystem1996,brownGPSFailureDetection1986,leeNewImprovedRAIM2006,hwangNIORAIMIntegrityMonitoring2005}. To detect multiple faults, all possible fault models have to be considered resulting in combinatorial complexity and heavy computational burden \cite{grovesPortfolioApproachNLOS2013,wangMultiConstellationGNSSPerformance2012,blanchOptimizedMultipleHypothesis2007,choiDemonstrationsMulticonstellationAdvanced2011,martiniReceiverIntegrityMonitoring2006}.

References to integrity monitoring in urban environment can be characterized as {\em measurement rejection} and {\em error characterization} methods. The first approach rejects faulty measurements to achieve consistent residuals and works well in open sky environment. The second approach identifies measurements contaminated by a large bias and assigns them corresponding weights but does reject them completely \cite{cosmen-schortmannIntegrityUrbanRoad2008}. 

An alternative approach  based on fusion of information from multiple sources (GNSS, IMU, map, odometry etc.) is elaborated. Previously published map-enabled approaches are described in e.g. \cite{toledo-moreoLaneLevelIntegrityProvision2010,velagaMapAidedIntegrityMonitoring2012}. The core idea of the presented algorithm is to evaluate the accuracy of pseudorange measurements based on the predicted tram position fixed to rail  and provide realistic variance and bias estimates for all individual satellites. Note that under NLOS conditions, multiple outliers from blocked sky segment may produce a large position error with small residuals, so the analysis based on residuals only \cite{castaldoPRANSACIntegrityMonitoring2014,salosReceiverAutonomousIntegrity2014} will not be efficient.

With linear complexity, this approach covers multiple fault detection with reasonable computational burden. Also -- unlike in the classical RAIM approach -- individual pseudoranges are characterized not only by the weights, but also the biases. Using Gaussian mixture distribution concept, the probability distribution function (pdf) of raw data and estimated biased pseudorange are approximated by a new pdf with equivalent first and second moments. In this way, biased measurements will strongly affect the resulting GNSS-based position estimate covariance ellipsoid which then serves as an input to a master fusion algorithm to calculate maximum a posteriori probability estimate of the tram position. 

The master fusion algorithm comes in the form of non-linear least squares problem utilizing the assumption of Gaussian distribution and conditional independence of the relevant sources of information. To see this, consider the Bayesian update
\begin{equation}
    p(x|a,b,c)\propto p(a,b,c|x)\ p(x),
    \label{eq:MAP}
\end{equation}
where $p(x) \sim \mathcal{N}(\hat{x},P_x)$ is the prior information and $a=g_a(x)\!+\!e_a,b=g_b(x)\!+\!e_b,c=g_c(x)\!+\!e_c$ are individual sources of information about $x$ with $\mbox{cov}\{e_{\bullet}\}=P_{\bullet}$. Under the aforementioned assumptions, the likelihood $l(x|a,b,c)=p(a,b,c|x)$ can be factorized into
\begin{equation}
    p(a,b,c|x)=p(a|x)p(b|x)p(c|x).
\end{equation}
Finding a maximum a posteriori (MAP) estimate that
maximizes \eqref{eq:MAP} for a given prior $\hat{x}$ and data $a, b, c$
\begin{multline}
    \hat{x}_{\rm MAP} = \argmax_x \exp{\left( -\|x\!-\!\hat{x}\|^2_{P^{-1}_x} -\|a\!-\!g_a(x)\|^2_{P^{-1}_a} \right.} \\ 
    -\left.\|b\!-\!g_b(x)\|^2_{P^{-1}_b} -\|(c\!-\!g_c(x)\|^2_{P^{-1}_c}\right),
\end{multline}
is equivalent to the non-linear least squares problem
\begin{multline}
    \min_x \left( \|x-\hat{x}\|^2_{P^{-1}_x} + \|a-g_a(x)\|^2_{P^{-1}_a}+ \right.\\
    \left. +\|b-g_b(x)\|^2_{P^{-1}_b}+\|c-g_c(x)\|^2_{P^{-1}_c}\right).
    \label{eq:NNLS}
\end{multline}
Solution of \eqref{eq:NNLS} using the Gauss-Newton method leads to the {\em iterated (extended) Kalman filter} algorithm.

\begin{figure}[t]
    \centering
    \includegraphics[width=0.69\linewidth]{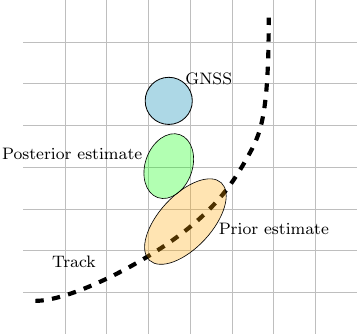}
    \caption{Illustration of the filtering step of the localization scheme
    without taking the inconsistencies of the individual pseudoranges
    into account. The coloured areas represent covariance ellipsoids of
    the specified quantities.}
    \label{fig:estimates_without_fancy}
\end{figure}

\begin{figure}[t]
    \centering
    \includegraphics[width=0.65\linewidth]{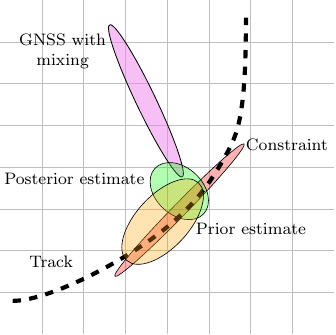}
    \caption{Illustration of the filtering step of the localization 
    scheme while taking inconsistencies of the individual pseudoranges 
    and track constraint into account. 
    The coloured areas represent covariance ellipsoids of
    the specified quantities.}
    \label{fig:fancy_estimates}
\end{figure}

The master fusion algorithm that is able to
incorporate GNSS pseudorange measurement inconsistencies and
the fact that the motion is constrained onto the track network directly
into the localization scheme can be illustrated graphically. 
Whereas the usual data update step of a
Kalman filter (see Figure \ref{fig:estimates_without_fancy}) combines only the 
position measurement (blue) with the prior
state estimate (orange), the
presented approach (see Figure \ref{fig:fancy_estimates}) processes the individual 
pseudorange measurements 
to form an inconsistency-aware reading (purple) and also forms
a soft constraint on the posterior state in the form of an artificial
position measurement in the vicinity of the track network (pink).
The posterior estimate lies closer to the tracks and the
covariance reflects the fact that the uncertainty of pseudorange
measurements coming from GNSS satellites in the segment of the sky
which is obscured by an obstacle (e.g. a building) is larger than
of the readings from other directions.

The paper is organized as follows: First, the iterated extended
Kalman filter framework is briefly introduced in section \ref{sec:inf_fusion}, along with its reformulation to an instance of a Gauss-Newton problem and how the constraint of the
state to the tram track network is constructed. In section 
\ref{sec:gnss_positioning} the basic principles of GNSS localization and
description of the pseudorange measurement data reconciliation
are described. In section \ref{sec:example} the application of the
proposed framework on a simulated and real data is presented.
The final part of the paper consists of a conclusion and possible
future improvements.

\section{INFORMATION FUSION}
\label{sec:inf_fusion}
State estimation seeks to find an estimate at time $k$, $\hat{\mathbf{x}}_k\in\mathbb{R}^n$, 
of the true state $\mathbf{x}_k$ at time $k$. For the specific case of 
iterated extended Kalman filter (IEKF), we assume that the system evolves
according to the discrete model in the form
% \begin{subequations}
\begin{align}
    \mathbf{x}_{k+1} &= \mathbf{f}_k(\mathbf{x}_k) + \mathbf{w}_k,\label{eq:state_equation}\\
    \mathbf{z}_k &= \mathbf{g}_k(\mathbf{x}_k) + \mathbf{v}_k,\label{eq:output_equation}
\end{align}
% \end{subequations}
where $\mbf{f}_k$ is the transition function, $\mbf{g}_k$ is the output function,
$\mbf{z}_k\in\mathbb{R}^m$ is the measurement vector,
$\mbf{w}_k\in\mathbb{R}^n$ and $\mbf{v}_k\in\mathbb{R}^m$ are the
process and measurement noises, respectively, which are not
correlated with the state vector nor between themselves.
Furthermore, the noise vectors are assumed to be white and normally
distributed around the zero vector of appropriate dimension with a given 
covariance matrix:
\begin{align}
    \mbf{w}_k &\sim \mathcal{N}(\mbf{0},\mbf{\Sigma}^{\mbf{w}}_k),\\
    \mbf{v}_k &\sim \mathcal{N}(\mbf{0}, \mbf{\Sigma}^{\mbf{v}}_k).
\end{align}
The IEKF algorithm consist of repeated computation of the filtering step
and the prediction step after the filter is initialized at the 
initial time step, $k=0$, with an initial estimate of the state and 
its covariance, i.e.
\begin{align}
    \hat{\mbf{x}}_{0|-1} &= \hat{\mbf{x}}_0,\\
    \mbf{P}_{0|-1} &= \mbf{P}_0.
\end{align}
The filtering step performs an inner loop of repeated linearization of the output
equation \eqref{eq:output_equation} around the current best estimate
of the state, $\hat{\mbf{x}}_k^i$ while utilizing the most recent
measurement $\mbf{z}_k$ (the upper index $i$ denotes the $i$-th iteration of the inner loop). The inner loop is initialized with
\begin{equation}
    \hat{\mbf{x}}_k^0 = \hat{\mbf{x}}_{k|k-1}.
\end{equation}
The following equations form the inner linearization loop:
\begin{gather}
    \mbf{G}_k^i = \left. \frac{\partial \mbf{g}_k}{\partial \mbf{x}_k} \right|_{\hat{\mbf{x}}_k^i},\label{eq:output_linearization}\\
    \mbf{K}_k^i = \mbf{P}_{k|k-1} \left(\mbf{G}_k^i\right)^T
        \left( \mbf{G}_k^i \mbf{P}_{k|k-1} \left(\mbf{G}_k^i\right)^T + 
            \mbf{\Sigma}_k^{\mbf{v}}\right)^{-1}, \label{eq:kalman_gain}\\
    \hat{\mbf{x}}_k^{i+1} = \hat{\mbf{x}}_{k|k-1} + \mbf{K}_k^i \left(
        \mbf{z}_k - \mbf{g}_k(\hat{\mbf{x}}_k^i) - \mbf{G}_k^i(\hat{\mbf{x}}_{k|k-1}-\hat{\mbf{x}}_k^i)\right).\label{eq:iekf_iteration}
\end{gather}
The computations \eqref{eq:output_linearization}-\eqref{eq:iekf_iteration} are performed
repeatedly with incrementing $i:=i+1$ until the subsequent iterations
fulfils the convergence condition 
$\|\hat{\mbf{x}}_k^i - \hat{\mbf{x}}_k^{i+1}\|<\varepsilon$
for a small predefined parameter $\varepsilon>0$. 
The posterior state estimate and its covariance are given as
\begin{align}
    \hat{\mbf{x}}_{k|k} &= \hat{\mbf{x}}_k^i, \\
    \mbf{P}_{k|k} &= (\mbf{I} - \mbf{K}_k^i\mbf{G}_k^i)\mbf{P}_{k|k-1}.
\end{align}

The prediction step consists of linearization of the state equation 
\eqref{eq:state_equation} at the current best estimate of the state and
propagating the covariance and state estimate through the state equation:
\begin{gather}
    \mbf{F}_k = \left. \frac{\partial \mbf{f}_k}{\partial\mbf{x}_k}\right|_{\hat{\mbf{x}}_k^i},\\
    \mbf{P}_{k+1|k} = \mbf{F}_k\mbf{P}_{k|k}\mbf{F}_k^T + \mbf{\Sigma}_k^\mbf{w},\\
    \hat{\mbf{x}}_{k+1|k} = \mbf{f}_k(\hat{\mbf{x}}_{k|k}).
\end{gather}
The filtering and prediction steps are computed for $k=1, 2,\ldots$ repeatedly in
alternating manner.

For the purpose of the next sections, it is useful to explicitly introduce
a way how to include additional information (readings from secondary sensors or 
other problem-specific data) into the IEKF framework. This would correspond to having
more than one output equation \eqref{eq:output_equation}. Consider (without loss
of generality) that we have an additional prior information we would like
to include in the filtering step, i.e. we can write
\begin{equation}
    \mbf{y}_k = \mbf{h}_k(\mbf{x}_k)+\mbf{u}_k
\end{equation}
for some function $\mbf{h}_k$ and normal noise $\mbf{u}_k\sim\mathcal{N}(\mbf{0},
\mbf{\Sigma}_k^\mbf{u})$. Under the assumption that $\mbf{y}_k$ are independent from
$\mbf{z}_k$, we can use the IEKF framework described above with augmented 
measurement vector, measurement noise covariances and the output function:
\begin{gather}
    \mbf{z}_k := \begin{bmatrix}
        \mbf{z}_k^T & \mbf{y}_k^T
    \end{bmatrix}^T, \label{eq:augment_z}\\
    \mbf{\Sigma_k^\mbf{v}} := \begin{bmatrix}
        \mbf{\Sigma}_k^\mbf{v} & \mbf{O}\\ \mbf{O} & \mbf{\Sigma}_k^\mbf{u}
    \end{bmatrix}, \label{eq:augment_sigma}\\
    \mbf{g}_k(\mbf{x}_k) := \begin{bmatrix}
        \mbf{g}_k^T(\mbf{x}_k) & \mbf{h}_k^T(\mbf{x}_k)
    \end{bmatrix}^T,\label{eq:augment_g}
\end{gather}
where $\mbf{O}$ is a zero matrix of appropriate dimension. If there are
more sources of additional information, the quantities above are augmented
with all of the required terms. This way, the soft constraint, GNSS reading and
the prior estimate from Figure \ref{fig:fancy_estimates} can be incorporated to
get the posterior estimate.

\subsection{Improving Filtering Step Convergence}
Repeating the procedure from \cite{fantaTramLocalizationUsing2024} which
itself is built upon the approaches presented in 
\cite{havlikPerformanceEvaluationIterated2015} and 
\cite{skoglundExtendedKalmanFilter2015}, one can interpret the filtering
step as an instance of a non-linear least squares problem.
Under the assumption that the prior distribution of the state
estimate (from the prediction step)\footnote{The symbol $\mbf{x}^k$ 
denotes the set of all historical values 
$\lbrace \mbf{x}_1,\ldots,\mbf{x}_k \rbrace$.}, $p(\mbf{x}_k|\mbf{z}^{k-1})$, and the likelihood of the measurement, $p(\mbf{z}_k|\mbf{x}_k)$, 
are normally distributed, we can write the following for the posterior
distribution
\begin{multline}
    p(\mbf{x}_k|\mbf{z}^k) \propto p(\mbf{z}_k|\mbf{x}_k)p(\mbf{x}_k|\mbf{z}^{k-1}) \\
    \propto \exp\left[ -\frac12 \left( [\mbf{z}_k-\mbf{h}_k(\mbf{x}_k)]^T(\mbf{\Sigma}_k^\mbf{v})^{-1}[\cdots] \right. \right. \\
    \left.\left. +[\hat{\mbf{x}}_{k|k-1}-\mbf{x}_k]^T\mbf{P}_{k|k-1}^{-1}[\cdots]\right) \right],
    \label{eq:posterior}
\end{multline}
where $[\cdots]$ denotes repetition of the same term in the last set
of brackets. To obtain the maximum a posteriori (MAP) estimate
we have find the maximizer of the exponential function \eqref{eq:posterior}.
Finding the posterior estimate is a result of the non-linear least squares optimization problem
\begin{multline}
    \argmin_{\mbf{x}_k\in\mathbb{R}^n} \left( [\mbf{z}_k-\mbf{g}_k(\mbf{x}_k)]^T\left(
    \mbf{\Sigma}_k^\mbf{v}\right)^{-1}[\cdots] \right.+\\ \left.[\hat{\mbf{x}}_{k|k-1}-\mbf{x}_k]^T\mbf{P}^{-1}_{k|k-1}[\cdots]\right) =\\
    \argmin_{\mbf{x}_k\in\mathbb{R}^n} V(\mbf{x}_k),
    \label{eq:criterion_reformulation}
\end{multline}
which can be solved using the Gauss-Newton method as described in
\cite{skoglundExtendedKalmanFilter2015}.
Rewriting \eqref{eq:iekf_iteration} into a recurrent
form (\cite{skoglundExtendedKalmanFilter2015}) results in
\begin{equation}
    \hat{\mbf{x}}_k^{i+1} = \hat{\mbf{x}}_k^i + \Delta_k^i,
    \label{eq:xhat_recurrent}
\end{equation}
where the step direction $\Delta_k^i$ is defined as
\begin{equation}
    \Delta_k^i = \hat{\mbf{x}}_{k|k-1}-\hat{\mbf{x}}_k^i + \mbf{K}_k^i\left(
        \mbf{z}_k - \mbf{g}_k(\hat{\mbf{x}}_k^i) - \mbf{G}_k^i(\hat{\mbf{x}}_{k|k-1}-\hat{\mbf{x}}_k^i)\right).
        \label{eq:delta}
\end{equation}
The reformulations in the \eqref{eq:criterion_reformulation} and
\eqref{eq:xhat_recurrent} allow us to improve convergence properties
of the filtering step. The basic IEKF scheme does not ensure
improvement of \eqref{eq:criterion_reformulation} in subsequent iterations
\begin{equation}
    V(\hat{\mbf{x}}_k^{i+1}) \leq V(\hat{\mbf{x}}_k^i),
    \label{eq:V_improvement}
\end{equation}
but using the reformulations above, we can perform a constrained line search
\begin{equation}
    \alpha^i = \argmin_{\alpha\in[0,1]} V(\hat{\mbf{x}}_k^i+\alpha\Delta_k^i),
\end{equation}
and perform the step of the iteration as
\begin{equation}
    \hat{\mbf{x}}_k^{i+1} = \hat{\mbf{x}}_k^i + \alpha^i\Delta_k^i
\end{equation}
to make sure \eqref{eq:V_improvement} does hold. The choice
of $\alpha = 1$ correspond to the usual EKF step. See 
\cite{havlikPerformanceEvaluationIterated2015,skoglundExtendedKalmanFilter2015}
for more details.

\subsection{Constraining the State Estimate}
To improve performance of a localization scheme for rail vehicles,
a heuristic approach to attract the state estimates towards tracks
was developed in \cite{fantaTramLocalizationUsing2024}. 
If tracks are modelled as a set of waypoints (which is often the case),
constraining the state estimate to the track network can be difficult
using traditional techniques described in e.g.\cite{simonOptimalStateEstimation2006} because
of the non-convex nature of such a constraint. 

The aforementioned heuristic technique fabricates a new position measurement
which always lies in the vicinity of tracks. When such measurement is incorporated into the
filtering scheme according to \eqref{eq:augment_z}-\eqref{eq:augment_g}, the
posterior estimate is driven closer to the track network. This soft constraint
is constructed from waypoints of the track network which lie close to the
prior estimate. Further details are described in \cite{fantaTramLocalizationUsing2024}.
The heuristic introduces additional hyperparameters which have to be tuned for
performance improvement.

\begin{figure}[t]
    \centering
    \includegraphics[width=0.85\linewidth]{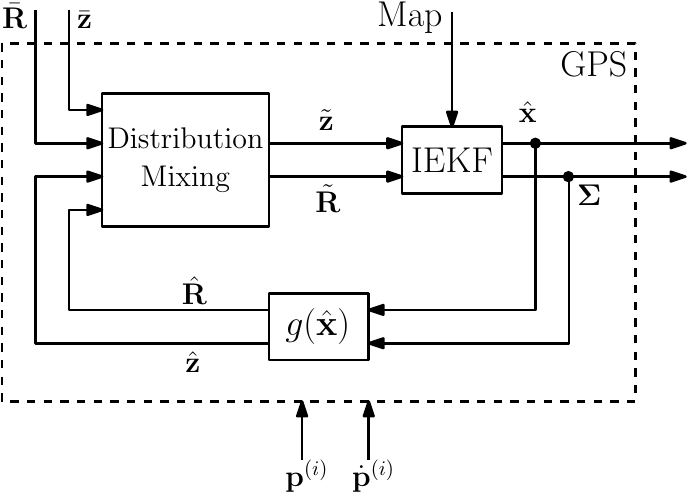}
    \caption{Proposed filtering scheme utilizing GNSS measurements and track map data.}
    \label{fig:diagram}
\end{figure}

%START OF JAKUB's PART, FEEL FREE TO CHANGE, MOVE, DELETE, WHATEVER...
\section{GNSS Positioning}
\label{sec:gnss_positioning}
GNSS satellites transmit a known pseudorandom code modulated onto the carrier signal. The receiver measures the 
frequency of the received carrier signal and compares it with its nominal value. The difference is the Doppler measurement which serves as a measurement of relative motion between the satellite and the receiver.

The receiver also cross-correlates the received code with the code 
transmitted by the satellite. The estimated delay serves as a measure of distance between the satellite and the receiver. The distance is corrupted by multiple sources of error. Two major sources of error in urban environment are the multipath effect where the receiver receives multiple replica of the signal and the non-line-of-sight (NLOS) error, where the receiver only receives a reflection of the signal from other objects.

The proposed localization scheme is shown in Figure \ref{fig:diagram}.
When a new vector of pseudoranges from the individual GNSS satellites,
$\bar{\mbf{z}}$, with the covariance $\bar{\mbf{R}}$ becomes available,
it is processed into a mixed distribution with the predicted measurement value ($\hat{\mbf{z}}=g(\hat{\mbf{x}})$, with covariance $\hat{\mbf{R}}$) corresponding to the current state estimate, $\hat{\mbf{x}}$ with covariance $\mbf{\Sigma}$ (the details are described below). The processed
GNSS readings, $\tilde{\mbf{z}}$ with covariance $\tilde{\mbf{R}}$, are passed into an IEKF which also takes the 
track map into the account when yielding the posterior state estimate 
(described above). The GNSS measurement is naturally influenced
by the position and velocity of the satellites, $\mbf{p}^{(i)}$ and $\dot{\mbf{p}}^{(i)}$, respectively.

%A multipath/NLOS error mitigation scheme is proposed in this paper which is based on using a mixed distribution of the measured pseudorange value and the value predicted by the IEKF is presented. The mitigation scheme is depicted in Figure \ref{fig:diagram} and will be discussed in detail in section \ref{sec:mixer}.

In this section, GNSS positioning solution is described. First GNSS pseudorange and Doppler observables models are described. Then a Kalman filter model for position estimation is presented. 
Lastly, equations for conversion of coordinates from Earth-centered, Earth-fixed (ECEF) frame to the geodetic (latitude, longitude, altitude) frame and to the east, north, up (ENU) frame are described.
\subsection{GNSS Observables Models}
Pseudorange and Doppler shift observables were used for position and velocity estimation. 
\subsubsection{Pseudorange}
The pseudorange measurement $\rho^{(i)}$ belonging to the $i$-th satellite can be modelled as
\begin{equation}
    \label{eq:pseudorange_model}
    \rho^{(i)} = r^{(i)}  + c (\delta t_r - \delta t^{(i)}) + I^{(i)} + T^{(i)} + M^{(i)} + \epsilon^{(i)},  
\end{equation}
where $r^{(i)}$ is the true distance from the receiver position at signal reception time to satellite position at transmission time, $c$ is the speed of light, $\delta t_r$ is the receiver clock offset, $\delta t^{(i)}$ is the satellite clock error, $I^{(i)}$ and $T^{(i)}$ are ionospheric and tropospheric delays, respectively, $M^{(i)}$ is the multipath error and $\epsilon^{(i)}$ are other unmodelled sources of error.

Receiver clock offset $\delta t_r$ needs to be estimated. The satellite clock error $\delta t^{(i)}$ is the sum 
\begin{equation}
     \delta t^{(i)} = \Delta_{\mathrm{\mathrm{clk}}} t^{(i)} + \Delta_{\mathrm{rel}} t^{(i)}.   
\end{equation}
The symbol $\Delta_{\mathrm{clk}} t^{(i)}$ denotes the satellite internal clock offset and can be computed using the parameters included in satellite navigation message \cite{noauthor_navstar_nodate} %TODO: ask: should I cite this?
and $\Delta_{\mathrm{rel}} t^{(i)}$ is relativistic clock error and can be computed as \cite{ashby_relativity_2003}
\begin{equation}
    \Delta_{\mathrm{rel}} = -2\frac{\mathbf{p}^{(i)} \cdot \mathbf{\dot{p}}^{(i)}}{c^2},
\end{equation} 
with $\mathbf{p}^{(i)}$, $\mathbf{\dot{p}}^{(i)}$ being the satellites position and velocity in ECEF coordinate system.

Different models exist to estimate $I^{(i)}$. Single frequency receivers rely on approximation models such as the
GPS Klobuchar model or Galileo NeQuick model (\cite{klobuchar_ionospheric_1987,nava_new_2008}) whose parameters are broadcasted in satellite navigation messages. The $I^{(i)}$ is frequency-dependent, so when using multi-frequency receiver, it can be compensated for using the equation from \cite{farrell_aided_2008}
\begin{equation}
    \hat\rho^{(i)} = \frac{f_1^2 \hat\rho_1^{(i)} - f_2^2 \hat\rho_2^{(i)}}{f_1^2 - f_2^2}.
\end{equation}
Here, $\hat\rho^{(i)}$ represents the ionosphere-delay-free pseudorange measurement, $f_1, f_2$ are frequencies of the carrier signals, and $\hat\rho_1^{(i)}, \hat\rho_2^{(i)}$ are their respective pseudorange measurements corrected for other known sources of error.

The tropospheric delay was neglected for the purposes of this paper, but models exist to compensate for the error \cite{ma_influence_2021,collins_assessment_1999,schuler_tropgrid2_2014}. We believe this simplification to be justified as the contribution of the tropospheric delay to the pseudorange measurement error is relatively low compared to other sources of error in urban environment such as the multipath error.

$M^{(i)}$ and $\epsilon^{(i)}$ were modelled as single additive white noise. Covariance and mean value of the noise were determined by mixing the distribution of the measured pseudorange and the theoretic pseudorange value distribution predicted by Kalman filter. This will be further discussed in section \ref{sec:mixer}.

\subsubsection{Doppler shift}
The Doppler shift observable is a measure of the relative motion between the receiver and the $i$-th satellite.  It can be modelled as
\begin{equation}
    \label{eq:doppler_model}
    \dot\rho^{(i)} = \lambda \Delta f = (\dot{\mathbf{p}}^{(i)} - \dot{\mathbf{p}}_r) \mathbf{L}^{(i)} + c \delta \dot{t}_r + \nu^{(i)},
\end{equation}
where $\dot\rho^{(i)}$ is the measurement of the magnitude of relative velocity, $\lambda$ is the carrier wavelength, $\Delta f$ is the measured deviation from carrier nominal frequency, $\mathbf{\dot p}^{(i)}, \mathbf{\dot p}_r$ are the true satellite and receiver velocities in ECEF frame, $\mathbf{L}^{(i)}$ is the line of sight vector
\begin{equation}
    \mathbf{L}^{(i)} = \frac{\mathbf{p}_r - \mathbf{p}^{(i)}}{\|\mathbf{p}_r - \mathbf{p}^{(i)}\|}, \label{eq:los}
\end{equation}
$\delta \dot t_r$ is receiver clock bias rate which needs to be estimated and $\nu^{(i)}$ is measurement noise assumed to be white and Gaussian.

\subsection{GNSS Kalman Filter Model}
\subsubsection{System model}
An interated extended Kalman filter was used to estimate the augmented state vector $\mathbf{x} = \begin{bmatrix}
    \mathbf{p}_r^T & \dot{\mathbf{p}}_r^T & c\delta t_r & c\delta \dot t_r
\end{bmatrix}^T$. 
The constant velocity model augmented by the clock error model
\begin{gather}
    \mathbf{x}_{k+1} = \mathbf{F} \mathbf{x}_k = 
        \begin{bmatrix}
            \mathbf{I} & \Delta t \, \mathbf{I} & 0 & 0 \\
            0 & \mathbf{I} & 0 & 0 \\
            0 & 0 & 1 & \Delta t \\
            0 & 0 & 0 & 1 
        \end{bmatrix} \mathbf{x}_k + \mathbf{\omega}_k
\end{gather}
was used,
where $\mathbf{\omega}_k$ is the process noise, $\mathbf{I}$ is the identity matrix and $\Delta t$ is the sampling period.
Subtracting the known deterministic errors from equation \eqref{eq:pseudorange_model} gives us the pseudorange measurement equation
\begin{align}
    \label{eq:pr_meas}
    \mathbf{g}_k^{\rho^{(i)}} (\mathbf{x_k}) &= \rho^{(i)} + \delta t^{(i)} - I^{(i)} \notag \\
    &= \|\mathbf{p}_r - \mathbf{p}^{(i)}\|  + c \delta t_r + \epsilon + M^{(i)}.
\end{align}
Linearizing equation \eqref{eq:pr_meas} around current state yields the pseudorange measurement matrix
\begin{align}
    \mathbf{G}_k^{\rho^{(i)}} &= 
        \begin{bmatrix} 
            L_x^{(i)} & 
            L_y^{(i)} & 
            L_z^{(i)} &
            0 & 0 & 0 &
            1 & 0
        \end{bmatrix},
\end{align}
where $L_x^{(i)},L_y^{(i)},L_z^{(i)}$ are the components of the line of sight vector $\mathbf{L}^{(i)}$.

Similarly, the Doppler measurement given by equation \eqref{eq:doppler_model} can be linearized around the current velocity and clock error rate, which yields the Doppler measurement matrix:
\begin{equation}
    \mathbf{G}_k^{\dot \rho^{(i)}} = 
        \begin{bmatrix} 
            0 & 0 & 0 &
            L_x^{(i)} & 
            L_y^{(i)} & 
            L_z^{(i)} &
            0 & 1
        \end{bmatrix}.
\end{equation}

\subsubsection{Measurement Noise Computation} \label{sec:mixer}
The variance $\mathbf{\bar R}$ of the pseudorange measurement noise $\epsilon^{(i)}$ was selected using the model used in \cite{realini_gogps_2013,ng_improved_2020}:
\begin{equation}
    \mathbf{\bar R} = \frac{10^{-\frac{CN_o-T}{a}}\left((\frac{A}{10^{-\frac{F-T}{a}}}-1)\frac{CN_o-T}{F-T}+1\right)}{\sin^2 E}, \label{eq:weight}
\end{equation}
where $E$ is the elevation angle, $CN_o$ is the carrier signal to noise ratio, $T = 50,\ F = 10,\ A = 30,\ a = 40$ are empirical constants. 

As previously mentioned, however, measurements are often corrupted not only by zero mean noise, but also by multipath error, which has unknown mean and covariance. To partially compensate for this, the measurement $\bar z$ and its covariance $\mathbf{\bar R}$ were not used directly in the data step of the Kalman filter. Instead a mixed distribution $\tilde z$ of the measurement $\bar z$ and the theoretical (predicted) value $\hat z$ of the measurement is used.

This pulls the mean values of the measurements closer to their predicted values and inflates the covariance of measurements that are highly inconsistent with the predicted position thus smoothing sudden jumps in pseudorange measurements caused by multipath and NLOS.

The mixing is done individually for each satellite. The computation of the mixed distribution is represented by the \textit{Distribution mixing} block in Figure \ref{fig:diagram}. The mixed pseudorange measurement distribution is described by the mean $\tilde z$ and variance $\mathbf{\tilde R}$. These later serve as the measurement and its covariance and are computed using equations (for clarity, superscripts ${}^{(i)}$ are omitted)
\begin{gather} 
    \mu_a = \frac{\bar{\mbf{R}}_k^{-1}}{\bar{\mbf{R}}_k^{-1} + \hat{\mbf{R}}_k^{-1}},  \label{eq:mixer_start}\\
    \mu_b = 1-\mu_a,                                                        \\
    \tilde z_k = \mu_a \bar z_k + \mu_b \hat z_k,                                 \\
    \tilde{\mbf{R}}_k = \mu_a (\bar{\mbf{R}}_k + (\bar z_k - \tilde z_k)^2) + \mu_b (\hat{\mbf{R}}_k + (\hat z_k - \tilde z_k)^2), \label{eq:mixer_end}
\end{gather}
where $\bar z_k$ is the value measured by the receiver corrected for known deterministic errors, $\mu_a, \mu_b$ are the weights of the measurement and the prediction and $\hat z_k, \mathbf{\hat R}$ are the predicted values of the measurement and the variance of the prediction computed by projecting the covariance of the position estimate onto the line of sight vector
\begin{align}
    \hat z_k &= g_k(\mathbf{x}_k), \\
    \mathbf{\hat R} &= \mathbf{L}^T \mathbf{P}_{k|k-1} \mathbf{L}.
\end{align}

\subsection{Coordinate Frame Conversions}
The GNSS positioning described above was done in ECEF frame. In order to convert position to geodetic coordinates, one needs to solve the following equations
\begin{align}
    p_x &= (R_N+h)\cos \phi \cos \lambda,    \label{eq:coords_start}\\
    p_y &= (R_N+h)\cos \phi \sin \lambda,   \\
    p_z &= (R_N+h-e^2 R_N)\sin \phi,          \\
    R_N &= a/(1-e^2 \sin^2\phi)^\frac{1}{2}, \label{eq:coords_end}
\end{align}
where $\phi$ is geodetic latitude, $\lambda$ is geodetic longitude, $h$ is height above ellipsoid, $a=\SI{6378137}{\metre}$ is WGS-84 equatorial radius and $e^2=0.00669437999$ is WGS-84 squared ellipsoid eccentricity.

Algorithms for solving equations \eqref{eq:coords_start}-\eqref{eq:coords_end} can be found in \cite{zhu_conversion_1994}. 

The velocity and covariance matrix can be rotated to local tangent plane (ENU frame) coordinates using the rotation matrix from \cite{farrell_aided_2008}
\begin{equation}
    \mathbf{T} =    
        \begin{bmatrix}
            -\sin\lambda            & \cos\lambda               & 0 \\
            -\cos\lambda \sin\phi   & -\sin\lambda\sin\phi      &\cos\phi \\
            \cos\lambda\cos\phi     & \sin\lambda\cos\phi       &\sin\phi
        \end{bmatrix}.
\end{equation} 
ENU velocity $\mathbf{v}_{\mathrm{ENU}}$ can then be computed as
\begin{equation}
    \mathbf{v}_{\mathrm{ENU}} = \mathbf{T}\mathbf{ \dot p}_r.
\end{equation}

Rotation matrix $\mathbf{T}$ can be also used to rotate the position or velocity covariance matrix
from ECEF to ENU frame
\begin{equation}
    \Sigma_{\mathrm{ENU}} = \mathbf{T} \Sigma_{\mathrm{ECEF}} \mathbf{T}^T.
\end{equation}

%END OF JAKUB's PART

\section{EXAMPLE}
\label{sec:example}

The proposed algorithm was first tested in a simulation and then on real data collected in Prague.
\subsection{Simulation}

\begin{figure}[t]
    \centering
    \includegraphics[width=0.652\linewidth]{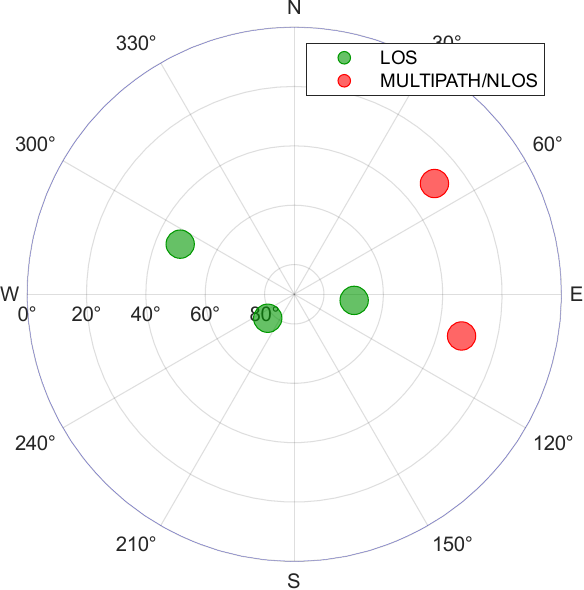}
    \caption{Skyplot of the visible satellites used in simulation. The green circles represent healthy measurements, red represent measurements corrupted by multipath or NLOS signals.}
    \label{fig:skyplot}
\end{figure}

A vehicle moving north in a straight line was simulated using a satellite constellation depicted in Figure \ref{fig:skyplot}. A total of 5 GPS satellites were simulated, 3 with clear line of sight (LOS) measurements during the entire simulation, 2 with NLOS measurements for a part of the simulation. The pseudorange measurements were simulated as the sum
\begin{equation}
    \bar\rho^{(i)} = \|{\mbf{p}_r-\mbf{p}^{(i)}}\| + \epsilon^{(i)} + b + M^{(i)},
\end{equation}
where $\bar\rho^{(i)}$ is the simulated measurement belonging to the \mbox{$i$-th} satellite, $\|{\mbf{p}_r-\mbf{p}^{(i)}}\|$ is the distance from the $i$-th satellite to the receiver, $b$ is a constant simulating the receiver clock bias, $\epsilon^{(i)} \sim \mathcal{N}(\SI{0}{\metre}, \SI{3}{\metre\squared})$ for LOS and $\epsilon^{(i)} \sim \mathcal{N}(\SI{0}{\metre}, \SI{10}{\metre\squared})$ for multipath/NLOS measurements is the measurement noise and $M^{(i)}=\SI{0}{\metre}$ for LOS and a uniformly distributed random value $M^{(i)} \subset (\SI{5}{\metre},\SI{20}{\metre})$ for multipath/NLOS measurements is the offset caused by the multipath/NLOS signal propagation.
Doppler measurements were not used in the simulation scenario.

\begin{figure}[t]
    \centering
    \includegraphics[width=0.85\linewidth]{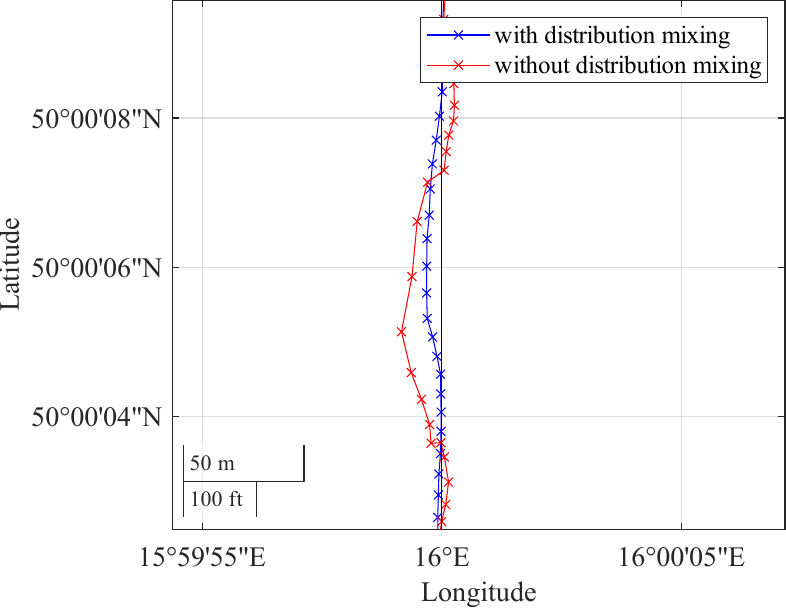}
    \caption{Illustration of the simulated effect of multipath on the position estimate before being projected onto the map of the track.}
    \label{fig:sim}
\end{figure}

An illustration of the simulated effect of the distribution mixing on the position estimate in multipath/NLOS environment is depicted in Figure~\ref{fig:sim}. The RMS error of the position estimate (RMS distance from ground truth) for different filters is shown in Table \ref{table:rms_gt_dist}. The baseline localization scheme which
does not utilize the mixed distribution or constraining the state
estimate results in the largest RMSE value. The RMSE lowers with
the proposed techniques (distribution mixing, constraining the state) being applied
during the estimation process.

\begin{table}[ht]
\centering
\caption{Root-mean-squared distance of the simulated position estimates from ground-truth}
\label{table:rms_gt_dist}
\begin{tabular}{l||c|c}
 & Without soft constr. & With soft constr. \\ \hline\hline
 With distribution mixing       &   $\SI{6.8}{\metre}$ & $\SI{6.0}{\metre}$ \\ \hline
 Without distribution mixing    &   $\SI{13.3}{\metre}$ & $\SI{9.7}{\metre}$ 
\end{tabular}
\end{table}

\subsection{Experiment}
The raw pseudorange and Doppler measurement data were collected using a U-Blox ZED-F9R satellite navigation receiver. The antenna was placed inside of a tram travelling along the route of line 22 in Prague. The data were recorded between the \textit{Vršovické náměstí} and \textit{Malostranská} stops. A~total of 1430 seconds with 1 second sampling period were recorded. The GPS, BeiDou and Galileo observables were used. OpenStreetMap data were used as an aiding map \cite{openstreetmapcontributorsPlanetDumpRetrieved2024}.

Perpendicular distances of the position estimates from the track according to the map were calculated. Table \ref{table:meandist} shows the root-mean-squared distance from the track map. 
For comparison, the RMS distance of state estimates provided by the U-Blox commercial solution was included.
As in Table~\ref{table:rms_gt_dist}, the proposed solution using
both the mixed distribution and the map aid results in the lowest
RMS distance value, however, without using the map aid, the filter
using the mixed distribution has higher RMS distance value than the
baseline filter. This is likely due to the fact that the distribution mixing has a low-pass effect on the measurement sequence. Therefore,
when the state is not constrained to the track network and
the vehicle remains under the effect of multipath/NLOS signals for a prolonged period of time, the position estimate slowly diverges from the true position. When the vehicle leaves the multipath/NLOS environment, the low-pass effect of the distribution mixing prevents
the state estimate from quickly recovering from the influence
of the past multipath/NLOS measurements. Constraining the state
estimate to the track network in each time step compensates this effect.

\begin{figure}[t]
    \centering
    \includegraphics[width=0.85\linewidth]{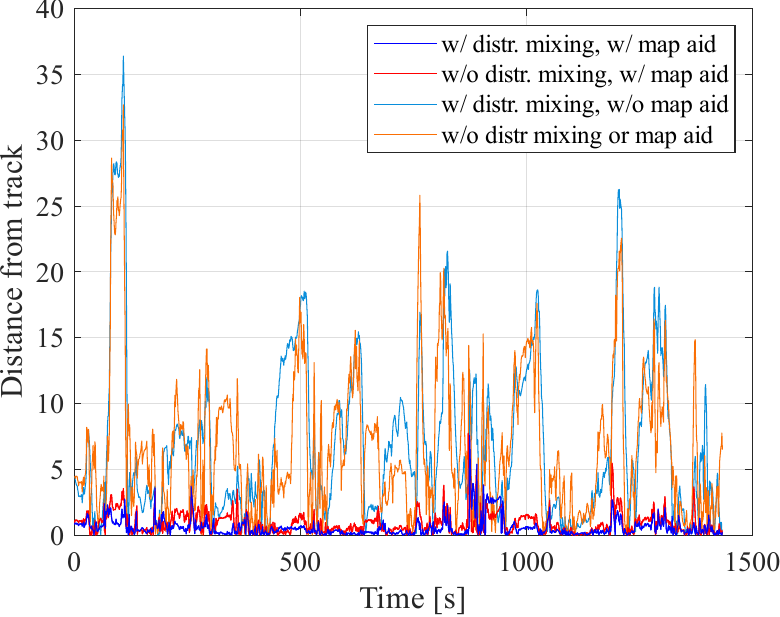}
    \caption{Distance of position estimates from the orthogonal projection onto track.}
    \label{fig:ecc_dist_from_track}
\end{figure}

Figure \ref{fig:ecc_dist_from_track} shows the development of the distance from track in time. It can be seen that the 
estimates coming from the filters which lack the constraint on the
state estimate lie further of the tram tracks for most of the
recorded period. 

\begin{figure}[t]
    \centering
    \includegraphics[width=0.85\linewidth]{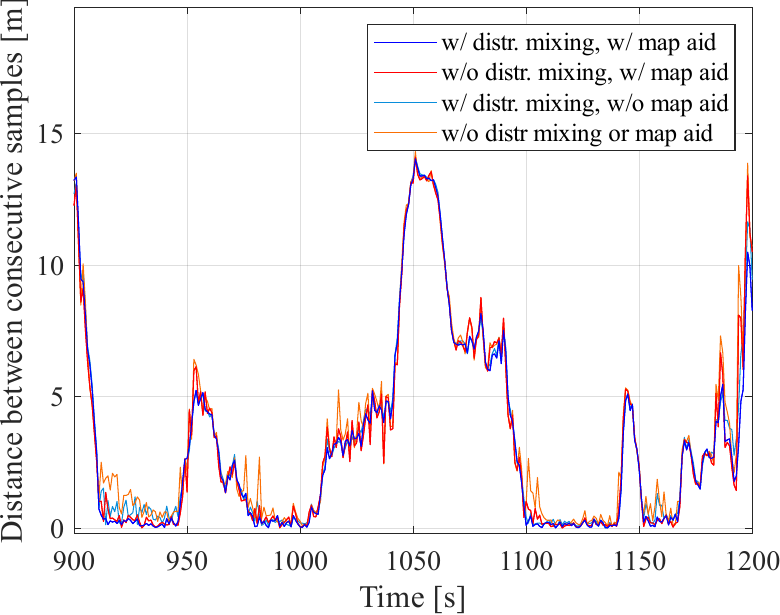}
    \caption{Distance between consecutive position estimates.}
    \label{fig:dist_samples}
\end{figure}

Figure \ref{fig:dist_samples} compares the distances between consecutive position estimates of the proposed filter that uses the distribution mixing and the baseline filter. As we can see, using distribution mixing leads to smoother trajectory that resembles realistic speed profile of the vehicle.

\begin{figure}[t]
    \centering
    \includegraphics[width=0.95\linewidth]{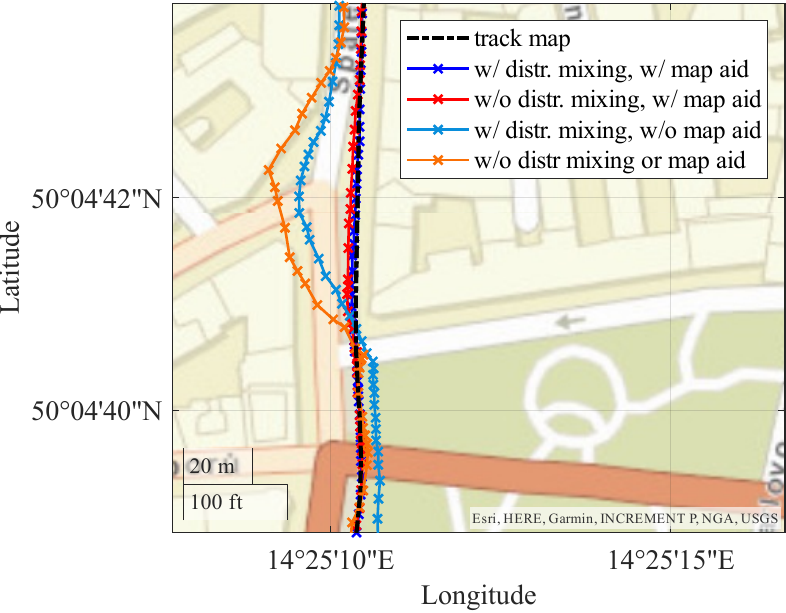}
    \caption{Visualized trajectories of the state estimates
    from the four filters. The accuracy deteriorates
    significantly when the trams enters an urban corridor.}
    \label{fig:ecc_karlak}
\end{figure}

Figure \ref{fig:ecc_karlak} shows the location estimates of the tested filters. The vehicle travelled northbound through a park (open space) where the position estimates matched the track. Then, as the vehicle entered a narrow street, 
the raw position measurements shifted west due to multipath-affected pseudorange measurements. The map aided estimates remained close to the track. The filter using the distribution mixing followed the track more closely than the baseline filter.


\begin{table}
\centering
\caption{Root-mean-squared distance of the position estimates from the orthogonal projection onto track map}
\label{table:meandist}
\begin{tabular}{l||c|c}
 & Without soft constr. & With soft constr. \\ \hline\hline
 With distribution mixing       & $\SI{9.29}{\metre}$ &   $\SI{0.91}{\metre}$ \\ \hline
 Without distribution mixing    & $\SI{8.55}{\metre}$ &   $\SI{1.19}{\metre}$ \\ \hline
 U-Blox solution                &         --              & $\SI{5.32}{\metre}$ 
\end{tabular}
\end{table}

\addtolength{\textheight}{-2.5cm}   % This command serves to balance the column lengths

\section{CONCLUSION}
A method for tram localization utilizing the combination of GNSS measurements with map of the track using soft-constrained iterated Kalman filter was presented. A GNSS pseudorange and Doppler measurement models together with a satellite measurement weighting scheme were described. 

The performance of the mixed distribution and the soft constraint based algorithm was tested in a simulation environment and also on real data. We have observed, that while using the mixed distribution, it is necessary to use map or other source of information to be able to avoid measurement drift. It was shown that activation of both of the features resulted
in best performance compared to the baseline filter which shows the potential of this approach for the problem of tram localization.

\subsection*{Future work}
    The presented filter might be further improved by using the interacting multiple models (IMM) method. This would allow us to better model different dynamic modes of tram movement such as stationary vehicle, vehicle moving in a straight line segment and a vehicle moving along a curve.
    This could, especially in the case of the stationary vehicle model, further decrease the drift of the measurement.
    Using bank of models would also allow us to extend the functionality to take track branching into account which would improve the positioning especially in tram depots where track-exact positioning is required but naive projection onto closest track might lead to errors.
    
    The proposed method of using the mixed distribution helps smooth out the estimates. However, when enough satellites are in view, it might be beneficial to discard some measurements completely or to not modify some at all. A decision scheme for outlier rejection should therefore be considered. Such scheme might employ $\chi^2$ based test to decide whether the measured pseudorange value falls within the predicted distribution or is biased. Also, dynamic bias tracking may be included to replace the presented "snapshot-based" approach. Finally, it could be beneficial to also constraint the state prediction step of the filter and not only during the data update step, which is currently the case.

%%%%%%%%%%%%%%%%%%%%%%%%%%%%%%%%%%%%%%%%%%%%%%%%%%%%%%%%%%%%%%%%%%%%%%%%%%%%%%%%

%%%%%%%%%%%%%%%%%%%%%%%%%%%%%%%%%%%%%%%%%%%%%%%%%%%%%%%%%%%%%%%%%%%%%%%%%%%%%%%%

%%%%%%%%%%%%%%%%%%%%%%%%%%%%%%%%%%%%%%%%%%%%%%%%%%%%%%%%%%%%%%%%%%%%%%%%%%%%%%%%

\section*{ACKNOWLEDGMENT}
Map data copyrighted OpenStreetMap contributors and available from \href{https://www.openstreetmap.org}{https://www.openstreetmap.org}.

\section*{REFERENCES}

\printbibliography[heading=none]

@article{realini_gogps_2013,
	title = {{goGPS}: open source software for enhancing the accuracy of low-cost receivers by single-frequency relative kinematic positioning},
	volume = {24},
	copyright = {http://iopscience.iop.org/info/page/text-and-data-mining},
	issn = {0957-0233, 1361-6501},
	shorttitle = {{goGPS}},
	url = {https://iopscience.iop.org/article/10.1088/0957-0233/24/11/115010},
	doi = {10.1088/0957-0233/24/11/115010},
	number = {11},
	urldate = {2024-10-26},
	journal = {Measurement Science and Technology},
	author = {Realini, Eugenio and Reguzzoni, Mirko},
	month = nov,
	year = {2013},
	pages = {115010},
}

@article{klobuchar_ionospheric_1987,
	title = {Ionospheric {Time}-{Delay} {Algorithm} for {Single}-{Frequency} {GPS} {Users}},
	volume = {AES-23},
	copyright = {https://ieeexplore.ieee.org/Xplorehelp/downloads/license-information/IEEE.html},
	issn = {0018-9251},
	url = {http://ieeexplore.ieee.org/document/4104345/},
	doi = {10.1109/TAES.1987.310829},
	number = {3},
	urldate = {2024-10-26},
	journal = {IEEE Transactions on Aerospace and Electronic Systems},
	author = {Klobuchar, John},
	month = may,
	year = {1987},
	pages = {325--331},
}

@article{ng_improved_2020,
	title = {Improved weighting scheme using consumer-level {GNSS} {L5}/{E5a}/{B2a} pseudorange measurements in the urban area},
	volume = {66},
	issn = {02731177},
	url = {https://linkinghub.elsevier.com/retrieve/pii/S027311772030404X},
	doi = {10.1016/j.asr.2020.06.002},
	language = {en},
	number = {7},
	urldate = {2024-10-26},
	journal = {Advances in Space Research},
	author = {Ng, Hoi-Fung and Zhang, Guohao and Yang, Kai-Yuan and Yang, Shi-Xian and Hsu, Li-Ta},
	month = oct,
	year = {2020},
	pages = {1647--1658},
}

@book{farrell_aided_2008,
	address = {New York},
	series = {Electronic engineering},
	title = {Aided navigation: {GPS} with high rate sensors},
	isbn = {978-0-07-149329-1},
	shorttitle = {Aided navigation},
	publisher = {McGraw-Hill},
	author = {Farrell, Jay},
	year = {2008},
	note = {OCLC: ocn212908814},
}

@misc{noauthor_navstar_nodate,
	title = {{NAVSTAR} {GPS} {Space} {Segment}/{Navigation} {User} {Segment} {Interfaces}},
    author = {Tony Anthony},
	url = {https://www.gps.gov/technical/icwg/IS-GPS-200N.pdf},
}

@article{zhu_conversion_1994,
	title = {Conversion of {Earth}-centered {Earth}-fixed coordinates to geodetic coordinates},
	volume = {30},
	copyright = {https://ieeexplore.ieee.org/Xplorehelp/downloads/license-information/IEEE.html},
	issn = {00189251},
	url = {http://ieeexplore.ieee.org/document/303772/},
	doi = {10.1109/7.303772},
	number = {3},
	urldate = {2024-10-29},
	journal = {IEEE Transactions on Aerospace and Electronic Systems},
	author = {Zhu, J.},
	month = jul,
	year = {1994},
	pages = {957--961},
}

@article{ma_influence_2021,
	title = {Influence of the inhomogeneous troposphere on {GNSS} positioning and integer ambiguity resolution},
	volume = {67},
	issn = {02731177},
	url = {https://linkinghub.elsevier.com/retrieve/pii/S0273117721000193},
	doi = {10.1016/j.asr.2020.12.043},
	language = {en},
	number = {6},
	urldate = {2024-10-31},
	journal = {Advances in Space Research},
	author = {Ma, Hongyang and Psychas, Dimitrios and Xing, Xuhuang and Zhao, Qile and Verhagen, Sandra and Liu, Xianglin},
	month = mar,
	year = {2021},
	pages = {1914--1928},
}

@book{collins_assessment_1999,
	series = {Technical report ({University} of {New} {Brunswick}. {Department} of {Geodesy} and {Geomatics} {Engineering})},
	title = {Assessment and {Development} of a {Tropospheric} {Delay} {Model} for {Aircraft} {Users} of the {Global} {Positioning} {System}},
	url = {https://books.google.cz/books?id=G2q7oAEACAAJ},
	publisher = {University of New Brunswick},
	author = {Collins, J.P.},
	year = {1999},
}

@article{schuler_tropgrid2_2014,
	title = {The {TropGrid2} standard tropospheric correction model},
	volume = {18},
	copyright = {http://www.springer.com/tdm},
	issn = {1080-5370, 1521-1886},
	url = {http://link.springer.com/10.1007/s10291-013-0316-x},
	doi = {10.1007/s10291-013-0316-x},
	language = {en},
	number = {1},
	urldate = {2024-10-31},
	journal = {GPS Solutions},
	author = {Schüler, Torben},
	month = jan,
	year = {2014},
	pages = {123--131},
}

@article{ashby_relativity_2003,
	title = {Relativity in the {Global} {Positioning} {System}},
	volume = {6},
	issn = {2367-3613, 1433-8351},
	url = {http://link.springer.com/10.12942/lrr-2003-1},
	doi = {10.12942/lrr-2003-1},
	language = {en},
	number = {1},
	urldate = {2024-10-31},
	journal = {Living Reviews in Relativity},
	author = {Ashby, Neil},
	month = dec,
	year = {2003},
	pages = {1},
}

@article{nava_new_2008,
	title = {A new version of the {NeQuick} ionosphere electron density model},
	volume = {70},
	copyright = {https://www.elsevier.com/tdm/userlicense/1.0/},
	issn = {13646826},
	url = {https://linkinghub.elsevier.com/retrieve/pii/S1364682608000357},
	doi = {10.1016/j.jastp.2008.01.015},
	language = {en},
	number = {15},
	urldate = {2024-10-31},
	journal = {Journal of Atmospheric and Solar-Terrestrial Physics},
	author = {Nava, B. and Coïsson, P. and Radicella, S.M.},
	month = dec,
	year = {2008},
	pages = {1856--1862},
}

@inproceedings{blanchOptimizedMultipleHypothesis2007,
  title = {An {{Optimized Multiple Hypothesis RAIM Algorithm}} for {{Vertical Guidance}}},
  author = {Blanch, Juan and Ene, Alex and Walter, Todd and Enge, Per},
  date = {2007-09-28},
  pages = {2924--2933},
  url = {http://www.ion.org/publications/abstract.cfm?jp=p&articleID=7644},
  urldate = {2024-11-04},
  eventtitle = {Proceedings of the 20th {{International Technical Meeting}} of the {{Satellite Division}} of {{The Institute}} of {{Navigation}} ({{ION GNSS}} 2007)},
  langid = {english}
}

@article{brownGPSFailureDetection1986,
  title = {{{GPS Failure Detection}} by {{Autonomous Means Within}} the {{Cockpit}}},
  author = {Brown, Grover and Hwang, Patrick Y. C.},
  date = {1986-12},
  journaltitle = {Navigation},
  volume = {33},
  number = {4},
  pages = {335--353},
  issn = {00281522},
  doi = {10.1002/j.2161-4296.1986.tb01485.x},
  url = {https://onlinelibrary.wiley.com/doi/10.1002/j.2161-4296.1986.tb01485.x},
  urldate = {2024-11-04},
  langid = {english}
}

@article{castaldoPRANSACIntegrityMonitoring2014,
  title = {P-{{RANSAC}}: {{An Integrity Monitoring Approach}} for {{GNSS Signal Degraded Scenario}}},
  shorttitle = {P-{{RANSAC}}},
  author = {Castaldo, Gaetano and Angrisano, Antonio and Gaglione, Salvatore and Troisi, Salvatore},
  date = {2014-09-23},
  journaltitle = {International Journal of Navigation and Observation},
  shortjournal = {International Journal of Navigation and Observation},
  volume = {2014},
  pages = {1--11},
  issn = {1687-5990, 1687-6008},
  doi = {10.1155/2014/173818},
  url = {https://www.hindawi.com/journals/ijno/2014/173818/},
  urldate = {2024-11-04},
  langid = {english}
}

@inproceedings{choiDemonstrationsMulticonstellationAdvanced2011,
  title = {Demonstrations of {{Multi-constellation Advanced RAIM}} for {{Vertical Guidance Using GPS}} and {{GLONASS Signals}}},
  author = {Choi, Myungjun and Blanch, Juan and Akos, Dennis and Heng, Liang and Gao, Grace and Walter, Todd and Enge, Per},
  date = {2011-09-23},
  pages = {3227--3234},
  url = {http://www.ion.org/publications/abstract.cfm?jp=p&articleID=9879},
  urldate = {2024-11-04},
  eventtitle = {Proceedings of the 24th {{International Technical Meeting}} of the {{Satellite Division}} of {{The Institute}} of {{Navigation}} ({{ION GNSS}} 2011)},
  langid = {english}
}

@inproceedings{cosmen-schortmannIntegrityUrbanRoad2008,
  title = {Integrity in Urban and Road Environments and Its Use in Liability Critical Applications},
  booktitle = {2008 {{IEEE}}/{{ION Position}}, {{Location}} and {{Navigation Symposium}}},
  author = {Cosmen-Schortmann, J. and Azaola-Saenz, M. and Martinez-Olague, M.A. and Toledo-Lopez, M.},
  date = {2008-05},
  pages = {972--983},
  issn = {2153-3598},
  doi = {10.1109/PLANS.2008.4570071},
  url = {https://ieeexplore.ieee.org/document/4570071},
  urldate = {2024-11-04},
  eventtitle = {2008 {{IEEE}}/{{ION Position}}, {{Location}} and {{Navigation Symposium}}}
}

@inproceedings{fantaTramLocalizationUsing2024,
  title = {Tram {{Localization}} Using {{Soft-Constrained Iterated Kalman Filter}} with {{Optimal Step Size Control}}},
  booktitle = {2024 {{IEEE International Conference}} on {{Multisensor Fusion}} and {{Integration}} for {{Intelligent Systems}} ({{MFI}})},
  author = {Fanta, Vít and Havlena, Vladimír and Hurák, Zdeněk},
  date = {2024-09-04},
  pages = {1--6},
  publisher = {IEEE},
  location = {Pilsen, Czech Republic},
  doi = {10.1109/MFI62651.2024.10705771},
  url = {https://ieeexplore.ieee.org/document/10705771/},
  urldate = {2024-10-24},
  eventtitle = {2024 {{IEEE International Conference}} on {{Multisensor Fusion}} and {{Integration}} for {{Intelligent Systems}} ({{MFI}})},
  isbn = {9798350368031}
}

@inproceedings{grovesPortfolioApproachNLOS2013,
  title = {A {{Portfolio Approach}} to {{NLOS}} and {{Multipath Mitigation}} in {{Dense Urban Areas}}},
  author = {Groves, P. D. and Jiang, Z. and Rudi, M. and Strode, P.},
  date = {2013-09-20},
  pages = {3231--3247},
  issn = {2331-5954},
  url = {http://www.ion.org/publications/abstract.cfm?jp=p&articleID=11264},
  urldate = {2024-11-04},
  eventtitle = {Proceedings of the 26th {{International Technical Meeting}} of the {{Satellite Division}} of {{The Institute}} of {{Navigation}} ({{ION GNSS}}+ 2013)},
  langid = {english}
}

@article{havlikPerformanceEvaluationIterated2015,
  title = {Performance Evaluation of Iterated Extended {{Kalman}} Filter with Variable Step-Length},
  author = {Havlík, Jindřich and Straka, Ondřej},
  date = {2015-11-19},
  journaltitle = {Journal of Physics: Conference Series},
  shortjournal = {J. Phys.: Conf. Ser.},
  volume = {659},
  pages = {012022},
  issn = {1742-6588, 1742-6596},
  doi = {10.1088/1742-6596/659/1/012022},
  url = {https://iopscience.iop.org/article/10.1088/1742-6596/659/1/012022}
}

@inproceedings{hwangNIORAIMIntegrityMonitoring2005,
  title = {{{NIORAIM Integrity Monitoring Performance In Simultaneous Two-Fault Satellite Scenarios}}},
  author = {Hwang, Patrick Y. and Brown, R. Grover},
  date = {2005-09-16},
  pages = {1760--1771},
  url = {http://www.ion.org/publications/abstract.cfm?jp=p&articleID=6377},
  urldate = {2024-11-04},
  eventtitle = {Proceedings of the 18th {{International Technical Meeting}} of the {{Satellite Division}} of {{The Institute}} of {{Navigation}} ({{ION GNSS}} 2005)},
  langid = {english}
}

@inproceedings{leeNewImprovedRAIM2006,
  title = {A {{New Improved RAIM Method Based}} on the {{Optimally Weighted Average Solution}} ({{OWAS}}) {{Under}} the {{Assumption}} of a {{Single Fault}}},
  author = {Lee, Y. C.},
  date = {2006-01-20},
  url = {https://www.semanticscholar.org/paper/A-New-Improved-RAIM-Method-Based-on-the-Optimally-a-Lee/c69af982f4aada4fe827b06d4901389388e33bb6},
  urldate = {2024-11-04}
}

@inproceedings{martiniReceiverIntegrityMonitoring2006,
  title = {Receiver {{Integrity Monitoring}} in {{Case}} of {{Multiple Failures}}},
  author = {Martini, I. and Wolf, R. and Hein, G. W.},
  date = {2006-09-29},
  pages = {2608--2620},
  url = {http://www.ion.org/publications/abstract.cfm?jp=p&articleID=6887},
  urldate = {2024-11-04},
  eventtitle = {Proceedings of the 19th {{International Technical Meeting}} of the {{Satellite Division}} of {{The Institute}} of {{Navigation}} ({{ION GNSS}} 2006)},
  langid = {english}
}

@misc{openstreetmapcontributorsPlanetDumpRetrieved2024,
  title = {Planet Dump Retrieved from {{https://planet.osm.org}}},
  namea = {OpenStreetMap contributors},
  nameatype = {collaborator},
  date = {2024},
  url = {https://www.openstreetmap.org/}
}

@book{parkinsonGlobalPositioningSystem1996,
  title = {Global {{Positioning System}}: {{Theory}} and {{Applications}}, {{Volume II}}},
  shorttitle = {Global {{Positioning System}}},
  editor = {Parkinson, Bradford W. and Enge, Per and Axelrad, Penina and Spilker Jr., James J.},
  date = {1996-01-01},
  publisher = {{American Institute of Aeronautics and Astronautics}},
  location = {Washington DC},
  doi = {10.2514/4.866395},
  url = {https://arc.aiaa.org/doi/book/10.2514/4.866395},
  urldate = {2024-11-04},
  isbn = {978-1-56347-107-0},
  langid = {english},
  pagetotal = {143-165}
}

@article{salosReceiverAutonomousIntegrity2014,
  title = {Receiver {{Autonomous Integrity Monitoring}} of {{GNSS Signals}} for {{Electronic Toll Collection}}},
  author = {Salós, Daniel and Martineau, Anaïs and Macabiau, Christophe and Bonhoure, Bernard and Kubrak, Damien},
  date = {2014-02},
  journaltitle = {IEEE Transactions on Intelligent Transportation Systems},
  volume = {15},
  number = {1},
  pages = {94--103},
  issn = {1558-0016},
  doi = {10.1109/TITS.2013.2273829},
  url = {https://ieeexplore.ieee.org/abstract/document/6578581},
  urldate = {2024-11-04},
  eventtitle = {{{IEEE Transactions}} on {{Intelligent Transportation Systems}}}
}

@book{simonOptimalStateEstimation2006,
  title = {Optimal State Estimation: {{Kalman}}, {{H}} [Infinity] and Nonlinear Approaches},
  shorttitle = {Optimal State Estimation},
  author = {Simon, Dan},
  date = {2006},
  publisher = {Wiley-Interscience},
  location = {Hoboken, N.J},
  isbn = {978-0-471-70858-2},
  pagetotal = {526}
}

@inproceedings{skoglundExtendedKalmanFilter2015,
  title = {Extended {{Kalman}} Filter Modifications Based on an Optimization View Point},
  booktitle = {2015 18th {{International Conference}} on {{Information Fusion}} ({{Fusion}})},
  author = {Skoglund, Martin A. and Hendeby, Gustaf and Axehill, Daniel},
  date = {2015-07},
  pages = {1856--1861},
  url = {https://ieeexplore.ieee.org/document/7266781},
  eventtitle = {2015 18th {{International Conference}} on {{Information Fusion}} ({{Fusion}})}
}

@article{toledo-moreoLaneLevelIntegrityProvision2010,
  title = {Lane-{{Level Integrity Provision}} for {{Navigation}} and {{Map Matching With GNSS}}, {{Dead Reckoning}}, and {{Enhanced Maps}}},
  author = {Toledo-Moreo, Rafael and Betaille, David and Peyret, François},
  date = {2010-03},
  journaltitle = {IEEE Transactions on Intelligent Transportation Systems},
  volume = {11},
  number = {1},
  pages = {100--112},
  issn = {1558-0016},
  doi = {10.1109/TITS.2009.2031625},
  url = {https://ieeexplore.ieee.org/abstract/document/5286855},
  urldate = {2024-11-04},
  eventtitle = {{{IEEE Transactions}} on {{Intelligent Transportation Systems}}}
}

@article{velagaMapAidedIntegrityMonitoring2012,
  title = {Map-{{Aided Integrity Monitoring}} of a {{Land Vehicle Navigation System}}},
  author = {Velaga, Nagendra R. and Quddus, Mohammed A. and Bristow, Abigail L. and Zheng, Yuheng},
  date = {2012-06},
  journaltitle = {IEEE Transactions on Intelligent Transportation Systems},
  volume = {13},
  number = {2},
  pages = {848--858},
  issn = {1558-0016},
  doi = {10.1109/TITS.2012.2187196},
  url = {https://ieeexplore.ieee.org/abstract/document/6162985},
  urldate = {2024-11-04},
  eventtitle = {{{IEEE Transactions}} on {{Intelligent Transportation Systems}}}
}

@article{wangMultiConstellationGNSSPerformance2012,
  title = {Multi-{{Constellation GNSS Performance Evaluation}} for {{Urban Canyons Using Large Virtual Reality City Models}}},
  author = {Wang, Lei and Groves, Paul D. and Ziebart, Marek K.},
  date = {2012-07},
  journaltitle = {The Journal of Navigation},
  volume = {65},
  number = {3},
  pages = {459--476},
  issn = {1469-7785, 0373-4633},
  doi = {10.1017/S0373463312000082},
  url = {https://www.cambridge.org/core/journals/journal-of-navigation/article/multiconstellation-gnss-performance-evaluation-for-urban-canyons-using-large-virtual-reality-city-models/7DCC8FC0C2F81222413E9A69FE409C24},
  urldate = {2024-11-04},
  langid = {english}
}

\end{document}